\documentclass[12pt]{article}

\newcommand{\beq}{\begin{equation}}
\newcommand{\eeq}{\end{equation}}
\newcommand{\ber}{\begin{array}}
\newcommand{\eer}{\end{array}}
\newcommand{\D}{{\cal D}}
\newcommand{\V}{{\cal V}}
\newcommand{\dtwo}{d^{\hspace{1pt}2}\hspace{-1pt}}
\newcommand{\del}{\partial}
\newcommand{\deln}{\partial_n}

\newcommand{\dsty}{\displaystyle}
\newcommand{\s}{\sigma}
\newcommand{\te}{\theta}

\newcommand{\de}{\delta}
\newcommand{\ds}{\dtwo\sigma}
\newcommand{\cnst}{\mbox{const}}

\newcommand{\eps}{\varepsilon}
\newcommand{\om}{\omega}

\begin{document}

\begin{titlepage}
\begin{flushright}
\end{flushright}
\vspace{15 mm}
\begin{center}
{\huge Worldline techniques for string theory solitons:\vspace{2mm}\\
recoil, annihilation and pair production.}
\end{center}
\vspace{12 mm}

\begin{center}
{\large
Oleg Evnin}\\
\vspace{3mm}
California Institute of Technology\\
Pasadena, CA 91125, USA
\end{center}
\vspace{20 mm}
\begin{center}
{\large Abstract}
\end{center}
\noindent

We analyze a model of interacting particles and strings described by a path integral with the Dirichlet boundary conditions. Such model is
a natural framework to examine the processes involving the center-of-mass motion
of string theory D0-branes: recoil, annihilation and pair production. We
demonstrate that, within the proposed formalism, the exclusive annihilation/pair-production amplitudes
admit a saddle point evaluation. Even though the saddle point equation
cannot be solved analytically, it allows to extract valuable information
on the coupling constant dependence of the amplitudes. In particular,
D0-brane pair production turns out to be suppressed as $\exp[-O(1/g_{st}^2)]$, much stronger than the
na\"\i ve expectation $\exp[-O(1/g_{st})]$. All our derivations generalize rather immediately
to the case of unstable D0-brane decay. In conclusion, we briefly comment
on the possible implications our results may have for the conventional
soliton-anti-soliton annihilation.

\vspace{1cm}
\begin{flushleft}
\today
\end{flushleft}
\end{titlepage}

\section*{Introduction}

The physics of topological defects in relativistic field systems is 
a rather accomplished area of study \cite{rajaraman}, yet fairly little 
remains known about how such localized objects are pair-produced in collisions
of elementary particles, and how they annihilate.
It is of very little use that the scattering of the fundamental fields quanta
off a topological defect (a process related to annihilation by crossing symmetry)
has been analyzed in quite some detail. The masses of the solitons are typically
inversely proportional to the coupling constant, forcing the annihilation products into the high energy regions of the phase space. Transition between the scattering and annihilation
kinematic regions would then require an analytic continuation {\it non-perturbatively} far in the phase space, and the {\it perturbative} expansion of the amplitude in the scattering region
will be essentially unrelated to the values of the amplitude in the annihilation region.

Even though there must exist a classical solution of the field equations describing the
annihilation process, it does not appear to be accessible to any degree of analytic control.
Indeed, there is apparently no straightforward way to prove that such a solution
encapsulates even the most fundamental features to be expected from the annihilation process, e.g.
the unitarity restrictions on the exclusive annihilation/pair-production amplitudes
that we'll discuss below.

One may expect that the situation becomes even more involved for the solitonic 
objects of string theory, the D-branes, since, in that case, we would have 
to deal with all the complexities of quantum gravity in addition to the problems
that plague the treatment of the field-theoretical process. It has been known since the seminal work 
by Polchinski \cite{inspiration} that the solitons of string theory admit
a particularly simple description in terms of the Dirichlet boundary 
conditions for open strings. Indeed, the analysis of elementary quanta 
scattering in the presence of D-branes is substantially simpler than the 
corresponding problem in the conventional relativistic field theory, as
it merely amounts to computing the expectation values of certain operators 
in a free 2-dimensional field system. One may therefore hope that 
worldsheet
techniques would provide a valuable tool for understanding the D-brane 
annihilation, a tool unavailable to the corresponding field theory
considerations.

Here, we attempt to construct a computational scheme for the D0-brane
(D-particle) annihilation amplitude inspired by the simplicity of the CFT
description of the extended D-branes. Since the center-of-mass motion of the D-particles
is of a crucial importance for the processes we intend to consider, it is only natural that we introduce an explicit functional integration over the D-particle worldlines, to which the string worldsheets
are attached via the Dirichlet boundary conditions. A similar method 
has been employed by Hirano and Kazama in their treatment of 
D-particle recoil \cite{hirano-kazama}. It has been shown that,
for low energy gravitational quanta, the scattering amplitudes obtained 
via this method can be matched with the standard no-recoil computation
and respect the consistency requirements, such as BRST-invariance.
Our goal in these notes is to address the question of how instructive
this kind of approach can be for essentially high-energy processes, 
in particular, the D-particle annihilation. Similarly to Hirano and Kazama's
paper, we restrict ourselves to bosonic D-particles, keeping
in mind that the generalization to the potentially physically interesting case of
superstring theory is likely to be conceptually straightforward, 
if only technically more challenging.

It is worth a notice that the attitude taken here is in many ways 
complementary to the attempts to address the same physical problem
within the tachyon condensation paradigm of string field theory
\cite{tachyon}. Indeed, the tachyon condensation approach typically deals with 
the initial state of coincident D-branes, whereas our aim is to analyze 
the genuine scattering\footnote{A scattering-like process in a tachyon condensation setting has been recently considered in \cite{hashimoto}.}. On the other hand, should there be a need to 
generalize the approach advocated here to the case of higher dimensional
D-branes, those will have to be wrapped around cycles in the compactification 
manifold, a requirement that does not appear within the string field theory
considerations.

Speaking of the possible phenomenological applications, one should certainly
mention the kinetics of primordial topological defects. Even though the
inflationary scenario suggests that the density of primordial topological
defects is dramatically diluted during the period of exponential expansion
\cite{vilenkin}, they could play an important role at the earlier stages
of the life of the Universe\footnote{In the cosmological context, the medium in which the topological defects move may exert substantial influence upon the annihilation process \cite{preskill}.}. The problem we're
considering is also in a (somewhat indirect) relation to the microscopic 
black hole pair production, a subject of much theoretical controversy
in the recent past \cite{voloshin}.

Determining the dependence of the annihilation amplitude on the
coupling constant is the central theme of our present investigation.
It is a requirement imposed by unitarity \cite{1/N, drukier} that
the {\it exclusive} annihilation amplitudes for the topological defects must 
be non-perturbatively suppressed, 
i.e. the amplitude should vanish identically when expanded in powers
of the coupling constant\footnote{Note that the {\it total} annihilation cross-section
does not need to be non-perturbatively suppressed, and it is expected to be comparable to the square of the size of the topological defect for the case of gauge theory. This is one of the indications that 
the final state multiplicity distribution for topological defect annihilation is typically rather non-trivial.}. The general intuition about how such suppression
could arise dynamically appeals to the well-known property that the inverse mass of the
topological defects is typically much smaller (by powers of the coupling
constant) than their size. Therefore, the typical wavelength of the 
annihilation products will be much smaller than the size of the
objects they are produced by, resulting in exponential damping.

The problem with this kind of explanations is that the path integral
includes arbitrarily singular configurations, and those could easily
upset the exponential damping, which is supposed to underlie the
non-perturbative suppression of the annihilation amplitude.
(The precise meaning of this statement will become more apparent once we
proceed with constructing the actual formalism.)
The saddle point estimate of the amplitude we obtain in the following sections
can be seen as a proof that the measure of those singular configurations
in the path integral is too small to alter the coupling constant dependence necessary to
maintain unitarity.

It is a fairly general principle that the processes involving topological
defects upset the decreasing significance of the final states with large multiplicity
familiar from the perturbative Feynman calculus. Producing a large number of quanta
may often turn out to be advantageous compared to the low multiplicity final states.
A possibly more familiar manifestation of this kind of behavior is the emission
of a large number of gluons in the instanton-induced processes of gauge theories
\cite{ringwald}. In our context, the bias for producing $O(1/g_{st})$ soft
quanta is even more dramatic, as we'll be able to see after completing
the saddle point evaluation of the amplitude.

\section*{Quantization of D-particle worldlines}

When dealing with extended D-branes, one does not have to worry much about 
the center-of-mass motion: the infinite D-brane mass makes recoil impossible,
at least when a localized object scatters off the extended D-brane. The 
situation is different, however, for compactified D-branes or genuine 
D-particles: for the low energy scattering experiments, the recoil is 
considerably suppressed by the mass of D-particle (inversely proportional 
to the coupling constant), but the center-of-mass motion becomes absolutely 
crucial in any essentially relativistic process, such as annihilation.

In the context of field theory, this problem has been addressed in 
\cite{christ}. One introduces the translational (center-of-mass) 
modes for the soliton explicitly, and performs quantization with the 
center-of-mass coordinate treated as a canonical variable. It has been
a subject of some controversy how the center-of-mass degrees of freedom
should be introduced for D0-branes \cite{fischler, tafjord}. Hirano and
Kazama's recipe \cite{hirano-kazama} constitutes a specific proposal to this 
end.

To account for the motion of the D-particle as a whole, one introduces its
coordinates explicitly and integrates over all the possible worldlines
$f^\mu(t)$, with $t$ being the proper time. The boundaries of the string 
worldsheet are restricted to the D-particle worldline, and the emission
of closed strings is described by insertions of the closed string vertex
operators in the interior of the worldsheet. Our main object of interest
is the amplitude for two D-particles starting off at the positions
$x_1^\mu$ and $x_2^\mu$ to annihilate into $m$ closed strings carrying
momenta $k_1$ to $k_m$:
\beq
\ber{l}
\dsty G(x_1,x_2|\,k_1,\cdots,k_m)=\sum\frac{\left(g_{st}\right)^\chi}{V_{CK}}
{\int[\D f]}_{\mbox{\small diff}}
\D t\,\D X\,\de\left(X_\mu(\te)-f_\mu(t(\te))\right)\vspace{2mm}\\
\dsty\hspace{4.5cm}\times\exp\left[-S_{D}(f)-S_{st}(X)\right]
\prod\limits_{a=1}^m\left\{g_{st}\V_a(k_a)\right\}
\eer
\label{master}
\eeq
where $S_D$ is the action for the D-particle to be discussed below,
$S_{st}$ is the standard (conformal gauge) Polyakov action
$$
S_{st}=\frac1{4\pi\alpha'}\int\ds\nabla X_\mu\nabla X^\mu,
$$
the integration with respect to $f_\mu$ extends over all the inequivalent
(unrelated by diffeomorphisms) curves starting at $x_1$ and ending at $x_2$,
the boundary of the worldsheet is parametrized by $\te$, and $t(\te)$ 
describes how this boundary is mapped onto the D-particle worldline.
The sum is over all the topologies of the worldsheets (not necessarily 
connected, but without any disconnected vacuum parts) and $\chi$ is 
the Euler number. $V_{CK}$ is the conformal Killing volume (the negative
regularized value of \cite{volume} should be used for the disk).
The fully integrated form of the vertex operators is implied.
We work in the Euclidean space-time, keeping in mind a subsequent analytic 
continuation to the Minkowski signature. The integration over moduli
of the worldsheet is suppressed for the rest of this paper, as
it does not affect the qualitative results. The annihilation amplitude 
can be deduced from (\ref{master}) by means of the standard reduction 
formula:
\beq
\ber{l}
\dsty\left<k_1,\cdots,k_m|p_1,p_2\right>\vspace{2mm}\\
\dsty\hspace{1.5cm}=\lim\limits_{p_1^2,p_2^2\to -M^2}
\left(p_1^2+M^2\right)\left(p_2^2+M^2\right)\int dx_1dx_2 e^{ip_1x_1} 
e^{ip_2x_2}G(x_1,x_2|\,k_1,\cdots,k_m) 
\eer
\label{reduct}
\eeq
where $M$ is the D-particle mass.

Several general questions have to be addressed regarding the expression
(\ref{master}). Firstly, the issue with the Weyl invariance appears to be 
rather subtle. On the physical grounds, one would believe that making the 
D-branes fully dynamical reinforces the consistency of the amplitudes, much 
in the same way as respecting the supergravity equations of motion 
makes the non-linear $\s$-models consistent. We shall examine below
how this works explicitly to the lowest order of the recoil perturbation
theory. The case of annihilation/pair-production is considerably less computationally
straightforward, and assessing the issues of the Weyl invariance there is
likely to require some new ideas.

Of course, whether or not the integration over the D-particle worldlines reinforces
the consistency of the string amplitudes depends crucially on the choice of the 
D-particle worldline action. It appears to be
a fairly general principle \cite{fradkin} that the value of the effective 
action for a background which couples to strings is given by (minus)
the sum of all connected vacuum string graphs evaluated in this 
background. Thus, very much in the spirit of \cite{combinatorics}:
\begin{equation}
S_D[f]=\sum\limits_{\mbox{connected}}{\frac{\left(g_{st}\right)^\chi}{-V_\chi}\int\D t\,\D X \,\de\left(X_\mu(\te)-f_\mu(t(\te))\right)
\exp\left[-S_{st}(X)\right]}
\label{SD}
\end{equation}
Again, the negative regularized value of the conformal Killing 
volume should be used for the disk \cite{volume}.
The exponentiation of the action in the path integral
can be seen as a result of summing up the disconnected graphs 
containing vacuum parts \cite{combinatorics}.
It can be shown that, for nearly straight 
worldlines, the above action is reduced to the na\"\i ve point-particle result
$M\int dt$. 
For curved worldlines, (\ref{SD}) would take into account the back-reaction 
from the space-time fields excited by the accelerating D-particle. Some
properties of this action will become more apparent as we proceed with the computation
of the amplitude.

Given the close relation between the integrand in (\ref{master}) and the worldline action
(\ref{SD}), it is convenient to rewrite (\ref{master}) in the following form:
\beq
\ber{l}
\dsty G(x_1,x_2|\,k_1,\cdots,k_m)=\sum\limits_{\mbox{all}}\frac{{\cal C}\left(g_{st}\right)^\chi}{V_{CK}}
{\int[\D f]}_{\mbox{\small diff}}
\D t\,\D X\,\de\left(X_\mu(\te)-f_\mu(t(\te))\right)\vspace{2mm}\\
\dsty\hspace{4.5cm}\times\exp\left[-S_{st}(X)\right]
\prod\limits_{a=1}^m\left\{g_{st}\V_a(k_a)\right\}
\eer
\label{all}
\eeq
Here, the sum extends over all the string diagrams including arbitrary disconnected vacuum
parts. There is no explicit action for the D-particle worldline, but it arises after resumming
the contributions from all the disconnected vacuum parts, which exponentiates to restore the $S_D$ of (\ref{SD}).
$\cal C$ is a combinatorial factor that can be deduced from (\ref{SD}).

Using the transformation properties of the vertex operators under the target space translations,
it is easy to see that
$$
G(x_1,x_2|\,k_i)=\exp\left[\frac{i}2(x_1^\mu+x_2^\mu)\sum k_i\right]
G\left(\left.\frac{x_1-x_2}2,-\frac{x_1-x_2}2\right|\,k_i\right)
$$
The first term here merely provides for the momentum conservation 
$\de$-function in the Fourier transform, and (\ref{reduct}) can be rewritten as
\beq
\ber{l}
\dsty\left<k_1,\cdots,k_m|p_1,p_2\right>=(2\pi)^{26}\de(p_1+p_2+\sum k_i)\vspace{2mm}\\
\dsty\hspace{2cm}\times\lim\limits_{p_1^2,p_2^2\to -M^2}
\left(p_1^2+M^2\right)\left(p_2^2+M^2\right)\int dx \exp\left[\frac{i}2(p_1-p_2)x\right] 
G\left(\left.\frac{x}2,-\frac{x}2\right|k_i\right) 
\eer
\label{reduct1}
\eeq
We shall work with this representation in our subsequent calculation of the amplitude.

\section*{The Gaussian integration}

By a direct inspection of (\ref{all}), it is easy to see that the integration 
over $X$ is Gaussian
and can be performed exactly. We will thus be able to recast the formalism
into a (0+1)-dimensional form. The Gaussian integration we have to perform
is closely related to the derivations in \cite{fradkin} and can be 
implemented by applying the formula:
\beq
\ber{l}
\dsty\int\D X\,\de\left(X(\te)-\xi(\te)\right)\,
\exp\left[\int \ds\left(-\frac1{4\pi\alpha'}\nabla X\nabla X+iJX\right)\right]\vspace{4mm}\\
\dsty\hspace{2cm}=\exp\left[-\pi\alpha'\int J(\s)D(\s,\s')J(\s')\ds\ds'-
i\int\xi(\te)\deln D(\te,\s')J(\s')d\te\ds'\right.\vspace{2mm}\\
\dsty\hspace{5cm}\left.
-\frac1{4\pi\alpha'}\int \xi(\te)\del_n\del_{n'}D(\te,\te')\xi(\te')d\te d\te'\right]
\eer
\label{gaussian}
\eeq
Here, $D$ is the Dirichlet Green function of the Laplace operator 
$\Delta D(\s,\s')=-\de(\s-\s')$, and $\del_n$
denotes the normal derivative evaluated at the boundary (which is parametrized
by $\theta$). It is convenient to consider
\beq
\ber{l}
\dsty G\left(\left.\frac{x}2,-\frac{x}2\right|J\right)=\sum\limits_{\mbox{all}}\frac{{\cal C}\left(g_{st}\right)^\chi}{V_{CK}}
{\int[\D f]}_{\mbox{\small diff}}
\D t\,\D X\,\de\left(X_\mu(\te)-f_\mu(t(\te))\right)\vspace{2mm}\\
\dsty\hspace{4.5cm}\times\exp\left[-S_{st}(X)+i\int \ds J_\mu X^\mu\right]
\eer
\label{master2}
\eeq
instead of (\ref{all}). Indeed, differentiating with respect to 
the source $J$ and setting it to $\sum k_i\de(\s-\s_i)$ allows us to reproduce 
the amplitude for an arbitrary vertex operator insertion. Performing the
integration in (\ref{master2}) by means of (\ref{gaussian}) yields:
\beq
\ber{l}
\dsty G\left(\left.\frac{x}2,-\frac{x}2\right|J\right)=\sum\limits_{\mbox{all}}\frac{{\cal C}\left(g_{st}\right)^\chi}{V_{CK}}
\exp\left[-\pi\alpha'\int J^\mu(\s)D(\s,\s')J_\mu(\s')\ds\ds'\right]\vspace{2.5mm}\\
\dsty\hspace{3cm}{\int[\D f]}_{\mbox{\small diff}}\,\D t\,
\exp\left[-i\int f^\mu(t(\te))\deln D(\te,\s')J_\mu(\s')d\te\ds'\right]\vspace{4mm}\\
\dsty\hspace{5cm}\exp\left[
-\frac1{4\pi\alpha'}\int f^\mu(t(\te))\del_n\del_{n'}D(\te,\te')f_\mu(t(\te'))
d\te d\te'\right]
\eer
\label{0+1}
\eeq
It should be noted that $D(\s,\s')$, $\chi$, $\cal C$ and $V$ depend on the topology
of the diagram corresponding to each particular term in the sum. For diagrams
with disconnected parts, $D(\s,\s')$ is block diagonal in the sense that it vanishes
whenever the two arguments belong to different disconnected components.

The path integral in (\ref{0+1}) may appear rather cumbersome, as one of 
the functions to be integrated over appears in the argument of the other one.
Nevertheless, the integration over $f^\mu(t)$ can be performed exactly by means
of a technique very similar to the treatment of the free point-like particle in
\cite{polyakov}.

We first rewrite the measure on reparametrization equivalence classes of $f^\mu(t)$
as
$$
[\D f]_{\mbox{\small diff}}=\D f\,\de\left[\dot{f}^2-1\right]=\D f \int\D z 
\exp\left[-\int\limits_0^Tz(\dot{f}^2-1)dt\right] 
$$
where $\de[\dot{f}^2-1]$ is a product of $\de$-functions at every point
(reinforcing $t$ to be the proper time), and, for each $t$, 
the integration over $z(t)$ is along a contour going from $c-i\infty$
to $c+i\infty$ in the complex $z$ plane, with $c$ being an arbitrary (positive) constant.
This contour can of course be deformed, an opportunity implicit in our subsequent
application of the saddle point method.

If we now introduce
$$
{\cal N}(t,t')=\int d\te d\te' \del_n\del_{n'}D(\te,\te')\, 
\de\left(t-t(\te)\right)\de\left(t'-t(\te')\right),\quad
d(t,\s)=\int d\te \deln D(\te,\s)\,\de\left(t-t(\te)\right)
$$
the $f$-integration in (\ref{0+1}) can be recast into a manifestly Gaussian
form:
$$
\ber{l}
\dsty G\left(\left.\frac{x}2,-\frac{x}2\right|J\right)=\sum\limits_{\mbox{all}}\frac{{\cal C}\left(g_{st}\right)^\chi}{V_{CK}}
\exp\left[-\pi\alpha'\int J^\mu(\s)D(\s,\s')J_\mu(\s')\ds\ds'\right]\vspace{2.5mm}\\
\dsty\hspace{3cm}\int\D t\,\D z\,\D f \exp\left[-\int z(\dot{f}^2-1)dt\right]
\exp\left[-i\int f^\mu(t)d(t,\s)J_\mu(\s)dt\ds'\right]\vspace{5mm}\\
\dsty\hspace{5cm}
\exp\left[-\frac1{4\pi\alpha'}\int f^\mu(t){\cal N}(t,t')f_\mu(t')dt dt'\right]
\eer
$$
It is convenient to pass on to the integration over $\D\dot{f}$ by means of the
relations
$$
\D f = dT\D\dot{f}\,\de\left(\int\limits_0^T\dot{f}^\mu dt+x^\mu\right)\qquad 
f^\mu(t)=\frac{x}2+\int\limits_0^t\dot{f}^\mu dt
$$
We should also keep in mind that
$$
\int\limits_0^T dt{\cal N}(t,t')=0\qquad\int\limits_0^T dt\,d(t,\s)=-1
$$
as
$$
\int d\te\,\del_n\del_{n'}D(\te,\te')=0\qquad\int d\te\,\deln D(\te,\s')=-1
$$
If we also perform the Fourier transform of (\ref{reduct1}), we arrive at the
following representation
$$
\ber{l}
\dsty G(p_1,p_2|J)=\int dx \exp\left[\frac{i}2(p_1-p_2)x\right] 
G\left(\left.\frac{x}2,-\frac{x}2\right|J\right)\vspace{3mm}\\
\dsty\hspace{1.7cm}=\sum\limits_{\mbox{all}}\frac{{\cal C}\left(g_{st}\right)^\chi}{V_{CK}}
\exp\left[-\pi\alpha'\int J^\mu(\s)D(\s,\s')J_\mu(\s')\ds\ds'\right]\vspace{2mm}\\
\dsty\hspace{3cm}\int dT \D t\,\D z\,\,e^{\int z\,dt}\int\D\dot{f} \,
\exp\left[i\int \dot{f}_\mu(t)\left(p_1^\mu-\int\limits_0^t
d(\tilde t,\s)J^\mu(\s)d\tilde t\ds\right)dt\right]\vspace{5mm}\\
\dsty\hspace{5cm}
\exp\left[-\int \dot{f}^\mu(t){\cal B}(t,t')\dot{f}_\mu(t')dt dt'\right]
\eer
$$
where we've introduced
$$
{\cal B}(t,t')=z(t)\,\de(t-t')+\frac1{4\pi\alpha'}\int\limits_0^t d\tilde t
\int\limits_0^{t'} d\tilde t'\,{\cal N}(\tilde t,\tilde t')
$$
At this point, the Gaussian integration becomes completely straightforward
and yields
\beq
\ber{l}
\dsty G(p_1,p_2|J)=\sum\limits_{\mbox{all}}\frac{{\cal C}\left(g_{st}\right)^\chi}{V_{CK}}
\exp\left[-\pi\alpha'\int J^\mu(\s)D(\s,\s')J_\mu(\s')\ds\ds'\right]\vspace{3mm}\\
\dsty\hspace{4cm}\int dT \D t(\te)\,\D z(t)\,\det[{\cal B}]^{-1/2}
\exp\left[\int z\,dt\right]\vspace{2mm}\\
\dsty
\exp\left[-\frac14\int \Big(p_1^\mu-\int\limits_0^t
d(\tilde t,\s)J^\mu(\s)d\tilde t\ds\Big)
{\cal B}^{-1}(t,t')\Big(p_{1\mu}-\int\limits_0^{t'}
d(\tilde t',\s')J_\mu(\s')d\tilde t'\ds'\Big)
dtdt'\right]
\eer
\label{zt}
\eeq
From this representation, it is apparent that the endpoints of the $z$ integration
contour can be moved towards $-\infty$. The contour itself cannot shrink to
$-\infty$, however, on account of the singularities of $(\det{\cal B})^{-1/2}$.
Should there be a discontinuity in $t(\te)$, these singularities move towards
$-\infty$ allowing the contour to be deformed arbitrarily far to the left in the
complex $z$ plane. Then the integrand will vanish due to the last factor in the
second line of (\ref{zt}). This is as it should be, since discontinuous
worldsheets do not give any contribution to the original path integral (\ref{master}).

The fact that we have been able to perform the integration over the D-particle
worldlines exactly is rather remarkable, since, for any {\it given} worldline, the
action $S_D$ featured in (\ref{master}) cannot be computed. This action however is itself constructed from string amplitudes and can be completely eliminated from the path integral 
at the cost of including the disconnected string worldsheets, as it has been done in (\ref{all}). This
peculiar simplification plays an important role in making possible the saddle point evaluation of the annihilation/pair-production amplitudes.

\section*{The saddle point method}

The presence of the $p_1$ and $J$ in the exponential of the last line of
(\ref{zt}) alludes to the relevance of the saddle point techniques, as those
quantities become non-perturbatively large in the annihilation kinematic
region. This can be made more apparent by introducing $\wp_i=p_i/M$ and ${\cal J}=J/M$
($M$ being the D-particle mass) and rewriting (\ref{zt}) as 
$$
\ber{l}
\dsty G(p_1,p_2|J)=\sum\limits_{\mbox{all}}\frac{{\cal C}\left(g_{st}\right)^\chi}{V_{CK}}
\exp\left[-\pi\alpha'M^2\int {\cal J}^\mu(\s)D(\s,\s'){\cal J}_\mu(\s')\ds\ds'\right]\vspace{3mm}\\
\dsty\hspace{4cm}\int dT\D t(\te)\,\D z(t)\,\det[{\cal B}]^{-1/2}
\exp\left[\int z\,dt\right]\vspace{2mm}\\
\dsty
\exp\left[-\frac{M^2}4\int \Big(\wp_1^\mu-\int\limits_0^t
d(\tilde t,\s){\cal J}^\mu(\s)d\tilde t\ds\Big)
{\cal B}^{-1}(t,t')\Big(\wp_{1\mu}-\int\limits_0^{t'}
d(\tilde t',\s'){\cal J}_\mu(\s')d\tilde t'\ds'\Big)
dtdt'\right]
\eer
$$
The latter representation is still somewhat inconvenient, since the position of
the saddle point depends on the saddle point parameter $M$. This dependence
is very simple, however, and can be completely eliminated by rescaling\footnote{
It should be kept in mind that, from now on, the geometrical parameters of
the worldsheet in physical space are related to those represented by
$t(\te)$ through a factor of $O(1/g_{st})$.}
$z(t)\to Mz(t/4\pi M\alpha')$, $t(\te)\to 4\pi M\alpha' t(\te)$, 
$T\to 4\pi M\alpha' T$, which brings the above integral to the form
\beq
\ber{l}
\dsty G(p_1,p_2|J)=\sum\limits_{\mbox{all}}\frac{{\cal C}\left(g_{st}\right)^\chi}{V_{CK}}
\exp\left[-\pi\alpha'M^2\int {\cal J}^\mu(\s)D(\s,\s'){\cal J}_\mu(\s')\ds\ds'\right]\vspace{3mm}\\
\dsty\hspace{4cm}\int dT \D t(\te)\,\D z(t)\,\det[{\cal B}]^{-1/2}
\exp\left[4\pi\alpha'M^2\int z\,dt\right]\vspace{2mm}\\
\dsty\hspace{1cm}
\exp\left[-\pi\alpha'M^2\int \Big(\wp_1^\mu-\int\limits_0^t
d\cdot {\cal J}^\mu d\tilde t\ds\Big)
{\cal A}^{-1}(t,t')\Big(\wp_{1\mu}-\int\limits_0^{t'}
d\cdot {\cal J}_\mu d\tilde t'\ds'\Big)
dtdt'\right]
\eer
\label{harness}
\eeq
where 
$$
{\cal A}(t,t')=z(t)\,\de(t-t')+\int\limits_0^t d\tilde t
\int\limits_0^{t'} d\tilde t'\,{\cal N}(\tilde t,\tilde t')
$$

In view of the subsequent application of the saddle point method, it is necessary
to specify how the integrand is analytically continued to the complex $t(\te)$.
It may seem worrisome that the definitions of ${\cal N}(t,t')$ and $d(t,\s)$
involve $\de$-functions, but these $\de$-functions appear in integral convolution,
and the result can be made manifestly analytic by expressing them through their
Fourier series on the interval $[0,T]$. Such a representation defines 
${\cal N}(t,t')$ and $d(t,\s)$ as periodic with respect to shifting $t$ or $t'$
by multiples of $T$. It is worth a notice that analogous analytic 
continuation is not possible at the level of (\ref{0+1}) due to the lack
of smoothness\footnote{It is a familiar fact (related to the central limit theorem) that the measure on random paths
is dominated by curves of fractal dimension 2: see, for example, \cite{ambjoern}.} 
in $f^\mu(t)$. In this respect, the summation over all the worldlines performed
in the previous section is a crucial prerequisite for the saddle point techniques
to be applicable.

With these specifications, the saddle point equations can be readily derived
by requiring the sum of the two exponents in (\ref{harness}) to be
stationary with respect to variations of $t(\te)$ and $z(t)$. In this manner,
one obtains
\beq
{\cal P}^\mu(t(\te))\left(\int d\te'\,\del_n\del_{n'}D(\te,\te')\int\limits_0^{t(\te')}
{\cal P}_\mu(t')dt'+\int\del_n D(\te,\s){\cal J}_\mu(\s)\ds\right)=0
\label{tsaddle}
\eeq
\beq
{\cal P}^\mu{\cal P}_\mu=-4
\label{zsaddle}
\eeq
where we've introduced
$$
{\cal P}^\mu(t)=\int\limits_0^T dt' {\cal A}^{-1}(t,t')
\left(\wp_1^\mu-\int\limits_0^{t'}d(\tilde t',\s'){\cal J}^\mu(\s')d\tilde t'\ds'\right)
$$
Multiplying (\ref{tsaddle}) by $\de(t-t(\te))$ and integrating with respect
to $\te$, we get
$$
\dot{z}\,{\cal P}^2+\frac12\,z\left({\cal P}^2\right)^\cdot=0
$$
which, coupled with (\ref{zsaddle}), implies $z=\cnst$. The relevant value of the constant
can be determined from the following argument. Because of the subsequent application of the
reduction formula, the value of the amplitude is determined by the divergent piece of
$G(p_1,p_2|J)$ as the momenta go on-shell. This divergence can only come from large values
of $T$ in the integral (which, of course, corresponds to the long distance propagation of
the on-shell particles). For large values of $T$, there will always be parts of the worldline
arbitrarily remote from the location of the worldsheet. There, (\ref{zsaddle}) will fix
$z=\pm i\sqrt{\wp^2}/2$ (with $\wp^2=\wp_1^2=\wp_2^2$). Since the saddle point value of $z$ does not depend on $t$,
we conclude that, for the purposes of evaluating the amplitude, $z$ can be set to
$\pm i\sqrt{\wp^2}/2$ everywhere.

Thus, we have two complex conjugate saddle points for $z$. After substituting the
corresponding values to (\ref{tsaddle}), we should obtain two complex conjugate
solutions for $t(\te)$. The equation (\ref{tsaddle}) is clearly beyond the reach of
analytic methods, but the saddle point configuration of the disconnected vacuum components
of the worldsheet can be easily identified as $t(\te)=\cnst$. Indeed, for $\te$ belonging
to a disconnected vacuum component, the second term in the brackets of (\ref{tsaddle})
vanishes. Enforcing $t(\te)=\cnst$ makes the first term in the brackets vanish as well, 
the equation being thereby satisfied.

Moreover, the values of the second functional derivative of the saddle point functional
with respect to $t(\te)$ can be calculated for the parts of the worldsheet which do not 
emit any final state particles. Indeed, if $\te$ and $\te'$ both belong to a disconnected
vacuum component with $t(\te)=t_0$,
$$
\ber{l}
\dsty\frac{\de^2}{\de t(\te)\de t(\te')}\left[\int \Big(\wp_1^\mu+\int\limits_0^t
d\cdot {\cal J}^\mu d\tilde t\ds\Big)
{\cal A}^{-1}(t,t')\Big(\wp_{1\mu}+\int\limits_0^{t'}
d\cdot {\cal J}_\mu d\tilde t'\ds'\Big)\right]\vspace{5mm}\\
\dsty\hspace{3cm}=-2{\cal P}^2(t_0)\del_n\del_{n'}D(\te,\te')=8\,\del_n\del_{n'}D(\te,\te')
\eer
$$
Furthermore, if $\te$ belongs to a disconnected vacuum component and $\te'$ belongs
to any other component of the worldsheet, the corresponding second derivative vanishes.

The above specifications in fact allow to perform a complete resummation 
(to the leading order of the saddle point approximation) of the contributions
to (\ref{harness}) coming from the disconnected vacuum components. Indeed, for each disconnected component in a given diagram, we'll obtain a factor
\beq
\int \D t(\te)\exp\left[-8\pi\alpha'M^2\int t(\te)\del_n\del_{n'}D(\te,\te') t(\te')d\te d\te'\right]
\label{vacuum}
\eeq
The value of this integral is a constant\footnote{One may argue that, for higher genus worldsheets, the modular integrations (that we do not write explicitly) will exhibit a tachyonic divergence. The same problem arises if one tries to compute the higher genus corrections to the D-particle mass. Both problems will disappear in the superstring case, which we see as a conceptual rationale behind the formal identification of the value of the integral (\ref{vacuum}) with the D-particle mass, which we're about to make.} (depending on the topology of each particular disconnected
vacuum component) times a factor of $T$, which comes from the integration over the
constant mode of $t(\te)$. If we now recall that the combinatorial coefficients $\cal C$
in (\ref{harness}) originated from expanding the exponential in (\ref{master}), it is easy to see that,
to the leading order of the saddle point approximation, the effect of all the 
disconnected vacuum components amounts to a factor of $\exp[\mu T]$, where $\mu$ is
a constant, which can
be evaluated by computing the Gaussian integral (\ref{vacuum}). In fact, it is easier to notice
that $G(p_1, p_2|J=0)$ is merely the D-particle propagator, and, identifying its pole with
the D-particle mass $M$, conclude that $\mu=-4\pi\alpha'M^2$.

For the parts of the worldsheet which do emit final state strings, it is not possible to
solve the equation (\ref{tsaddle}) explicitly. 
However, valuable information can be extracted from the mere assumption that
a saddle point exists
(this can be checked in the low energy scattering case).
It is an important circumstance that, for the worldsheets located not too close
to the endpoints of the worldline, in the limit $T\to\infty$, the value of the
saddle point function does not change under the shifts of $t(\te)$ by a constant
(we assume $\wp_1^2=\wp_2^2=\wp^2$, as it is on-shell).
It is not hard to take into account this quasi-zero mode though, as the subsequent
application of the reduction formula discards everything but the term growing
most rapidly as $T$ goes to $\infty$. Since extending the interval to which $t$ and $t'$
belong to $[-\infty,\infty]$ does not introduce any singularities to (\ref{tsaddle}),
we should expect that, as $T$ goes to $\infty$, the solutions to (\ref{tsaddle})
located not too close to the endpoints of the worldline approach a fixed shape.
The saddle point evaluation of the $\D t$ and $\D z$ integrals in (\ref{harness})
can be therefore written (up to the pre-exponential factors) as
\beq
e^{-4i\pi\alpha'M^2 T\sqrt{\wp^2}}\left(T e^{-\alpha'M^2{\cal F}[{\cal J}]} + o(T)\right)
\label{q0}
\eeq
where $\cal F$ does not depend on $T$ and can be determined from the solution to
the equation (\ref{tsaddle}) with $t$ and $t'$ running from $-\infty$ to $\infty$. 
The above expression is written for the saddle point $z=i\sqrt{\wp^2}/2$,
the contribution from $z=-i\sqrt{\wp^2}/2$ is the complex conjugate of this.
We shall omit the $o(T)$ term in the following, as it vanishes upon the
application of the reduction formula.

Assembling everything together, up to the pre-exponential factors, the saddle point
estimate of (\ref{harness}) is
$$
\ber{l}
\dsty \bigg[G(p_1,p_2|J)\bigg]_{\mbox{\small saddle}}\approx
\sum\frac{\left(g_{st}\right)^\chi}{V_{CK}}
\exp\left[-\pi\alpha'M^2\int {\cal J}^\mu(\s)D(\s,\s'){\cal J}_\mu(\s')\ds\ds'\right]\vspace{3mm}\\
\dsty\hspace{3cm}\times\,e^{-\alpha'M^2{\cal F}[{\cal J}]}\int\limits_0^\infty T 
\exp\left[-4\pi\alpha'M^2\left(1+i\sqrt{\wp^2}\right)T\right]dT + \mbox{c.c.}
\eer
$$
where the summation is once again performed only over those worldsheets which
do not contain any disconnected vacuum parts. Performing the $T$ integration and
applying the reduction formula, we finally obtain
\beq
\ber{l}
\dsty\left<p_1,p_2|J\right>_{\mbox{\small saddle}}\approx
\sum\frac{\left(g_{st}\right)^\chi}{V_{CK}}
\exp\left[-\pi\alpha'\int J^\mu(\s)D(\s,\s')J_\mu(\s')\ds\ds'\right]\vspace{3mm}\\
\dsty\hspace{5cm}
\times\left(\exp\left\{-\alpha'M^2{\cal F}[J/M]\right\}+ \mbox{c.c.}\right)
\eer
\label{saddle}
\eeq
To compute the actual amplitude, one has to combine this expression and its functional
derivatives with respect to $J$ so as to imitate the vertex operator insertions, as well as to integrate over
the worldsheet moduli and the positions of the vertex operators (these will
be saddle point integrations themselves, in analogy to \cite{gross}). All these tasks are likely to require
numerical computations, as is the evaluation of the $\cal F$-functional. 

An essential piece of information contained in the expression (\ref{saddle}) is, 
however, that it
identifies the dependence of the annihilation amplitude on the coupling constant.
We can see that, for a final state with the number of quanta fixed as $g_{st}\to 0$ and
momenta growing proportionally to the mass of the D-particle, the amplitude
diminishes as $\exp(-O(1/g_{st}^2))$. Indeed, the anticipated non-perturbative suppression
has become manifest\footnote{There doesn't appear to be any straightforward way to
prove that $\cal F$ is greater than zero in the entire annihilation kinematic
region (after the analytic continuation to Minkowski signature). Nevertheless, (\ref{saddle}) suggests that
the amplitude is either non-perturbatively suppressed or non-perturbatively enhanced,
and the inherent absurdity of the latter option is quite apparent.}. It is worth noting that the character of this coupling dependence is
not at all obvious at the intermediate stages of our computation. Only upon the
application of the saddle point method can we see that the qualitative arguments
outlined in the introduction do, in fact, result in a non-perturbative suppression
of the amplitude.

It may appear rather surprising that the estimate of the amplitude obtained above is exactly the same
as if we had used the simplistic free particle action $M\int dt$ in place of the
expression (\ref{SD}) which takes into account the emission and re-absorption of virtual
string states by the accelerating D-particles. The qualitative explanation here is
that, being objects of effective size $\sqrt{\alpha'}$, the D-particles are not
likely to emit {\it virtual} string states of energy much higher than $1/\sqrt{\alpha'}$. These
states will produce worldline curvatures of order $g_{st}$ and will not therefore
affect the leading order of the saddle point approximation. One may have suspected that the worldline curvature
could become large in the region where the {\it final} state strings are emitted. Yet,
the saddle point equation (\ref{tsaddle}) implies that the amplitude is dominated
by worldsheets of physical size $\sim 1/g_{st}$ (which corresponds to $t(\te)$ being of order 1). Therefore, the emission occurs from a segment
of the worldline whose size is of order $1/g_{st}$, and the relevant curvatures
are of order $g_{st}$ again.

\section*{Recoil perturbation theory and the Fischler-Susskind mechanism}

Let us now take a step back and examine how the more conventional coupling expansion of the recoil perturbation theory arises in our set-up. Namely, let's set aside the (necessarily relativistic) annihilation/pair-production processes and consider closed string states
scattering off a D-particle. Since the mass of the D-particle diverges in the limit
$g_{st}\to 0$, it can be treated as static to the lowest order in $g_{st}$ (if we keep
the momenta of the incident closed strings fixed as $g_{st}\to0$), and the corrections  
due to the motion of the D-particle's center-of-mass (i.e. recoil) will appear as
a perturbative expansion in powers of $g_{st}$. This is the familiar recoil perturbation
theory.

The formalism of the previous section is, of course, perfectly applicable in this case.
Since the momenta of the closed string states (and hence $J$) are kept fixed as $g_{st}\to 0$
(and hence $M\to\infty$), ${\cal J}\equiv J/M$ can be treated as perturbation. The saddle point
equation (\ref{tsaddle}) implies then that the saddle point configuration $t(\te)=\cnst+O(1/M)$, i.e. the path integral (\ref{harness}) is dominated by nearly
constant $t(\te)$ (the fluctuations of $t(\te)$ are of order $1/M$ generically, and so
is the saddle point value, as we've just remarked). One could look for solutions of the
saddle point equation as an expansion in ${\cal J}$. However, it is much more efficient to
generate the perturbation theory directly from the path integral (\ref{harness}) by
constructing a suitable expansion of the ``effective action'' functional of (\ref{harness})
$$
\begin{array}{l}
\dsty S_{eff}[z(t),t(\te)]=-4\pi\alpha'M^2\int z\,dt\vspace{2mm}\\
\dsty\hspace{1cm}+\pi\alpha'M^2\int \Big(\wp_1^\mu-\int\limits_0^t
d\cdot {\cal J}^\mu d\tilde t\ds\Big)
{\cal A}^{-1}(t,t')\Big(\wp_{1\mu}-\int\limits_0^{t'}
d\cdot {\cal J}_\mu d\tilde t'\ds'\Big)dtdt'
\end{array}
$$

There is one difficulty one encounters in implementing such a program.
It is a most straightforward approach to try to construct a Taylor-like expansion
of $S_{eff}$ in powers of $t(\te)$ around $t(\te)=\cnst$:
$$
\begin{array}{l}
\dsty S_{eff}(\cnst+t(\te))\vspace{2mm}\\
\dsty\hspace{1cm}=S_{eff}(\cnst)+\int d\te \left.\frac{\de S_{eff}}{\de t(\te)}\right|_{\mbox{\tiny const}}t(\te)+\frac12\int d\te d\te' \left.\frac{\de^2 S_{eff}}{\de t(\te)\de t(\te')}\right|_{\mbox{\tiny const}}t(\te)t(\te')+\cdots
\end{array}
$$
This strategy comes to mind in immediate relation to the computational techniques most commonly used in $\s$-models, 
and it has been employed in \cite{hirano-kazama} for the purposes we're presently pursuing here. Unfortunately, such an expansion does not exist. In \cite{hirano-kazama}, 
various $\zeta$-function prescriptions have been devised (in the lowest order) to deal with the infinities arising when one tries to brute-force the Taylor-like expansion, but the situation quickly becomes hopeless, if one tries to envisage the general structure of the
recoil perturbation theory in such a framework.

The origin of the above complication can be traced back to the
non-analytic properties of the worldlines in the path integral (\ref{0+1}).
Indeed, for non-analytic $f^\mu(t)$ (most worldlines are fractal and therefore
non-differentiable \cite{ambjoern}) the ``effective action'' in (\ref{0+1}) does not admit a
Taylor-like expansion in $t(\te)$ around {\it any} configuration of $t(\te)$.
The integration over $f^\mu(t)$ improves the situation considerably: the
resulting effective action can be expanded around any $t(\te)\ne\cnst$, 
but the non-analyticity still survives for the worldsheets whose boundary
shrinks to a single point. (One may become worried about whether such non-analyticity
could undermine the application of the saddle point method from the previous section.
Whereas any discussions of explicit contour deformation in multi-dimensional case
are necessarily rather subtle, there are still good chances that the deformation
implicit in the saddle point evaluation of the integral can be attained, since
$S_{eff}$ {\it is} analytic for any $t(\te)\ne\cnst$.)

Luckily, the Taylor-like expansion is not the only way to generate a sensible
perturbation theory. Appearing as insertions in a Gaussian path integral,
the exponentials of $t(\te)$ are just as tractable as powers of $t(\te)$.
We shall therefore resort to a combination of a Taylor-like and a Fourier-like
expansion. We first introduce 
$$
A(t,t')=i\frac{\sqrt{\wp^2}}{2}\de(t-t')\qquad B(t,t')=\de z(t)\,\de(t-t')+\int\limits_0^t d\tilde t
\int\limits_0^{t'} d\tilde t'\,{\cal N}(\tilde t,\tilde t')
$$ 
such that ${\cal A}(t,t')=A(t,t')+B(t,t')$, and expand formally
\begin{equation}
{\cal A}^{-1}=\frac1A-\frac1AB\frac1A+\frac1AB\frac1AB\frac1A+\cdots
\label{AB}
\end{equation}
(we work with one of the two saddle points in $z(t)$,
the other one will give a complex conjugate contribution, as in the previous section). Upon inserting this expression in the $S_{eff}$ we can isolate the following terms
that need to be retained in the exponent (everything else is small and may be treated
perturbatively, as it will become apparent after we exhibit the powers of the
expansion parameter $1/M$):
\beq
\begin{array}{c}
\dsty S_{eff}^{(1)}=4\pi i\alpha'M^2\sqrt{\wp^2}T\vspace{2mm}\\
\dsty S_{eff}^{(2)}=-4\pi\alpha'M^2\int dtdt'\int\limits_0^t d\tilde t
\int\limits_0^{t'} d\tilde t'\,{\cal N}(\tilde t,\tilde t')\equiv
-4\pi\alpha'M^2\int t(\te)\del_n\del_{n'}D(\te,\te')t(\te') d\te d\te'\vspace{2mm}\\
\dsty S_{eff}^{(3)}=-\frac{8\pi i\alpha'M^2}{\sqrt{\wp^2}}\int{\de z^2}dt\vspace{2mm}\\
\dsty S_{eff}^{(4)}=\frac{4\pi i\alpha'M^2}{\sqrt{\wp^2}}\wp_1^\mu\int dt\int\limits_0^t
d\cdot {\cal J}_\mu d\tilde t\ds\equiv\frac{4\pi i\alpha'M}{\sqrt{\wp^2}}\wp_1^\mu\int t(\te)\del_nD(\te,\s)
J(\s)_\mu d\te\ds\end{array}
\label{seffgauss}
\eeq
The first term here merely provides for the correct pole structure. It will disappear
after integration over $T$ and application of the reduction formula. The remaining three
terms define a Gaussian integral with respect to $z(t)$ and $t(\te)$.

A general term in the expansion (\ref{AB}) can be constructed as follows.
We should take a certain number of factors
\beq
\int\limits_0^{t_i} d\tilde t
\int\limits_0^{t_{i+1}} d\tilde t'\,{\cal N}(\tilde t,\tilde t')
\label{Ntt}
\eeq
and, for each $i$ add a factor 
$$
\frac{\left(\de z(t_i)\right)^{n_i}}{(i\sqrt{\wp^2}/2)^{-n_i+1}}
$$
where $n_i$ are non-negative integers. Then, for the first and last $t_i$ we either add
a factor of $\wp_1^\mu$ or a factor of
\beq
\int\limits_0^{t_i}
d(\tilde t,\s) {\cal J}^\mu (\s) d\tilde t\ds
\label{dt}
\eeq
We then integrate over all $t_i$'s and multiply the result by $\pi\alpha'M^2$.

We now proceed with the Gaussian integral over $\de z(t)$. As it can be seen from the expression for $S_{eff}^{(3)}$, each factor of $\de z(t)$ will produce a factor of $1/M$ in the result of the Gaussian integration. Where several $\de z$'s occur at the same $t_i$, divergencies from the singularity of the $\de z$ propagator (proportional to the $\de$-function) will be present. Just like in the case of free particle considered in \cite{polyakov}, these divergencies will merely renormalize
the saddle point (expectation) value of $z(t)$.

We are left with a Gaussian integral over $t(\te)$. As we've remarked before,
the structures of the type (\ref{Ntt}) and (\ref{dt}) cannot be expanded in powers
of $t(\te)$ around $t(\te)=\cnst$. Instead, we'll Fourier-trasform in $t_i$'s:
$$
\begin{array}{c}
\dsty\int\limits_0^{t_i} d\tilde t
\int\limits_0^{t_{i+1}} d\tilde t'\,{\cal N}(\tilde t,\tilde t')\to\frac1{\kappa_i\kappa_{i+1}}
\int e^{i\kappa_it(\te)}\del_n\del_n'D(\te,\te')e^{i\kappa_{i+1}t(\te')}d\te d\te'\vspace{3mm}\\
\dsty\int\limits_0^{t_i}
d(t,\s) {\cal J}^\mu (\s) d\tilde t\ds\to \frac1{\kappa_i}\int e^{i\kappa_it(\te)}\del_nD(\te,\s)
{\cal J}^\mu(\s)d\te\ds
\end{array}
$$
At this point, the integration over $t(\te)$ can be explicitly performed. The substitution
$t(\te)\to t(\te)/M$, $\kappa_i\to M\kappa_i$ reveals the additional powers of $1/M$ that each term
will receive upon the integration over $t(\te)$.

Having sketched the general structure of the recoil perturbation theory, we shall now turn
to the important issue of how the consistency of the string amplitudes is respected to the
lowest non-trivial order. As we've remarked early on in the course of our present investigation, the inclusion of curved D-particle worldlines into the path integral (\ref{master}) will generally induce a Weyl anomaly thereby threatening the decoupling
of the negative norm states, which is a crucial requirement for the consistency of the
string theory S-matrix. The educated hope is that (a suitable modification of) the familiar
Fischler-Susskind mechanism \cite{fischler-susskind,polchinski-fischler-susskind} will come to rescue the formalism that
has been constructed so far. Indeed, it is a well known fact that there are singularities in
the modular integration for the higher genus worldsheets coming from the corners of the moduli space. Should we have chosen the
D-particle worldline action correctly, the Weyl anomalies induced by those singularities
will precisely cancel the Weyl anomalies coming from the coupling to an accelerating
D-particle. Let us see how this works out to the first non-trivial order of the
recoil perturbation theory.

The Fischler-Susskind mechanism in our set-up implies a cancellation between an infrared
divergence on a higher genus worldsheet and an ultraviolet divergence on a lower genus
worldsheet (in the sense of cancellation of the corresponding Weyl anomalies).
To the lowest order, this means that we should examine the ultraviolet divergencies
on a disk to the order $1/M$ and compare them with the modular integration divergencies
from an annulus coupled to a straight D-particle worldline (the curved worldlines will only
contribute to the higher orders in $1/M$).

When we expand $S_{eff}$ (for a disk) according to (\ref{AB}) and treat all the terms
except for the ones indicated in (\ref{seffgauss}) as a perturbation, we can identify
the following insertions that can in principle contribute to the order $1/M$ in the 
resulting recoil perturbation theory Gaussian integral:
$$
\begin{array}{c}
\dsty F_1=-\frac{2\pi i\alpha'M^2}{\sqrt{\wp^2}}\int dt\left(\int\limits_0^{t}
d(\tilde t,\s) {\cal J}^\mu (\s) d\tilde t\ds\right)^2\vspace{2mm}\\
\dsty F_2=\frac{8\pi\alpha'M^2}{\wp^2}\wp^\mu\int dtdt'\int\limits_0^{t} d\tilde t
\int\limits_0^{t'} d\tilde t'\,{\cal N}(\tilde t,\tilde t')\int\limits_0^{t'}d\tilde t''\ds
d(\tilde t'',\s) {\cal J}^\mu (\s)\vspace{2mm}\\
\dsty F_3=\frac12\left(\frac{8\pi\alpha'M^2}{\wp^2}\wp^\mu\int dt \de z(t)\int\limits_0^{t}
d(\tilde t,\s) {\cal J}^\mu (\s) d\tilde t\ds\right)^2
\end{array}
$$
After we integrate over $\de z(t)$ and Fourier-transform, these three terms become:
$$
\begin{array}{c}
\dsty F_1\sim\alpha'\int \frac{d\kappa}{\kappa^2}\left|\int e^{i\kappa t(\te)}\del_nD(\te,\s)
J_\mu(\s)d\te\ds\right|^2\vspace{2mm}\\
\dsty F_2\sim\alpha'M\wp_1^\mu\int \frac{d\kappa}{\kappa^2}\int d\te d\te' t(\te)\del_n\del_n'D(\te,\te')e^{i\kappa t(\te')}\int e^{-i\kappa t(\te'')}\del_nD(\te'',\s)
J_\mu(\s)d\te''\ds\vspace{2mm}\\
\dsty F_3\sim\alpha'\int\frac{d\kappa}{\kappa^2}\left|\wp_1^\mu \int e^{i\kappa t(\te)}\del_nD(\te,\s)J_\mu(\s)d\te\ds\right|^2
\end{array}
$$
We now have to compute the path integral
$$
\int\D t(\te)\,\left(F_1+F_2+F_3\right)\exp\left[-S_{eff}^{(2)}-S_{eff}^{(4)}\right]
$$
The relevant structure that is featured in all the three terms is
$$
\begin{array}{l}
\dsty\int \frac{d\kappa}{\kappa^2}\int\D t(\tilde\te)\,e^{i\kappa\left(t(\te)-t(\te')\right)}\vspace{2mm}\\
\dsty\times\exp\left[-4\pi\alpha'M^2\int t(\tilde\te)\del_n\del_{n'}D(\tilde\te,\tilde\te')t(\tilde\te') d\tilde\te d\tilde\te'+\frac{4\pi i\alpha'M}{\sqrt{\wp^2}}\wp_1^\mu\int t(\tilde\te)\del_nD(\tilde\te,\s)
J_\mu(\s)d\tilde\te\ds\right]\vspace{3mm}\\
\dsty\hspace{.5cm}\sim\frac1{\alpha'M}\exp\left[-\pi\alpha'\wp_1^\mu\wp_1^\nu\int J_\mu(\s)\del_nD(\tilde\te,\s)[\del_n\del_{n'}D]^{-1}(\tilde\te,\tilde\te')\del_{n'}D(\tilde\te',\s')
J_\nu(\s')d\tilde\te d\tilde\te'\ds\ds'\right]\vspace{2mm}\\
\dsty\hspace{3cm}\times\int \frac{d\kappa}{\kappa^2}\exp\left[-2\kappa^2\log\eps+\kappa h_1(\te,\te')+\kappa^2h_2(\te,\te')\right]
\end{array}
$$
Where $\log\eps\equiv\del_n\del_{n'}D(\te,\te)$ is the regularized value of the
singularity of the boundary-to-boundary propagator ($\eps$ being the worldsheet cut-off),
and $h_{1,2}(\te,\te')$ are certain (cut-off independent) functions. If we now substitute $\tilde\kappa=\kappa/\sqrt{\log\eps}$, we observe that the functions $h_{1,2}$ do not contribute
to the UV divergent piece of the path integral\footnote{The IR divergence at $\kappa=0$ is not what interests us here, and it is in fact unphysical. It must disappear in a more accurate treatment taking into account that $t$ only varies within the finite range $[0,T]$.}, which turns out to be $(\te,\te')$-independent and proportional to
\beq
\frac{\sqrt{\log\eps}}{\alpha'M}\,e^{-\pi\alpha'\wp_1^\mu\wp_1^\nu\int J_\mu(\s)\del_{n\phantom{'}}\hspace{-1mm}D(\tilde\te,\s)[\del_{n\phantom{'}}\hspace{-1mm}\del_{n'}D]^{-1}(\tilde\te,\tilde\te')\del_{n'}D(\tilde\te',\s')
J_\nu(\s')d\tilde\te d\tilde\te'\ds\ds'}
\label{logeps}
\eeq
Of course, the $\kappa$ integral above is divergent for small {\it finite} values of $\eps$,
and the expression we've given must be understood in terms of analytic continuation in $\eps$.
As we'll see below, an analogous analytic continuation is needed for the annulus modular integration. This is very similar to what has been described in \cite{tafjord}.

Now, in the path integral of $F_2$, the expression (\ref{logeps}) will appear in convolution
with $\del_n\del_{n'}D(\te,\te')$, and, since the latter is orthogonal to constant modes,
the path integral of $F_2$ will not contain any UV-divergent piece. Both path integrals
of $F_1$ and $F_3$ will simplify due to the relation
$$
\int\del_nD(\te,\s)
J^\mu(\s)d\te\ds=-\int J^\mu\ds=p_1^\mu-p_2^\mu
$$
and the UV divergent part of the path integral of $F_3$ will turn out to be of order $1/M^2$,
not $1/M$ due to the relation $2\wp_1(\wp_1-\wp_2)=(\wp_1-\wp_2)^2$. If we also notice that
the exponential in (\ref{logeps}) is merely the value of the Gaussian integral
$$
\int\D t(\te)\,\exp\left[-S_{eff}^{(2)}-S_{eff}^{(4)}\right]
$$
(without any insertions), we conclude that the UV divergent part of the disk amplitude
to the order $1/M$ is proportional to
\beq
\frac{\left(p_1-p_2\right)^2}{\alpha'M}A_{D_2}^0\sqrt{\log\eps}
\label{D2}
\eeq
where $A_{D_2}^0$ is the zeroth order value of the disk amplitude, i.e. the amplitude
computed in the background of a straight D-particle worldline, in neglect of recoil.

We now have to compare this result with the divergence in the modular integration of the
annulus amplitude to the lowest order in $1/M$, i.e. the annulus amplitude in the background
of a straight D-particle worldline. The computation of such an annulus amplitude is fairly standard. The fact that the disk amplitude $A_{D_2}^0$ makes an appearance in the expression
(\ref{D2}) should be taken optimistically as far as the prospective cancellation of divergencies is concerned, since the familiar plumbing fixture construction of \cite{polchinski-fischler-susskind} relates the divergencies in the modular integration to amplitudes on worldsheets of lower genus. Let us see in some more detail how this actually works out.

Along the lines of \cite{polchinski-fischler-susskind}, the annulus amplitude with an insertion of the operators $V^{(1)},\cdots,V^{(n)}$ (in the interior) can be expressed through the disk amplitudes with additional operator insertions at the boundary as follows:
$$
\left<V^{(1)}\cdots V^{(n)}\right>_{\mbox{\small annulus}}=\sum\limits_\alpha \int dq q^{h_\alpha-2} \int d\te d\te' \left<V_\alpha(\te)V_\alpha(\te')V^{(1)}\cdots V^{(n)}\right>_{D_2}
$$
where the summation extends over a compete set of local operators $V_\alpha(\te)$ with
conformal weights $h_\alpha$, and $q$ is the gluing parameter that can be related to
the annular modulus. The divergence in the integral over $q$ coming from the region $q\approx0$
(corresponding to an annulus degenerating into a ring) will be dominated by the terms with
the smallest possible $h_\alpha$. Neglecting the tachyon divergence, which is a pathology
peculiar to the case of the bosonic string, we identify (in close relation to the investigations of \cite{fischler}) the following set of relevant operators (parametrized
by an integer $i\in[1,26]$ and a real $\om$):
$$
V^i(\te,\om)=:\del_n X^i(\te)e^{i\om X^0(\te)}:
$$
with conformal weights $h=1+\alpha'\om^2/4$. For small values of $q$ (which is the region
we're interested in) only small values of $\om$ will contribute into the integral.
We therefore transform the annular divergence as follows:
$$
\ber{l}
\dsty\left<V^{(1)}\cdots V^{(n)}\right>_{\mbox{\small annulus}}\sim \int dq d\om q^{-1+\alpha'\om^2/4}\int d\te d\te' \left<V^i(\te,\om)V^i(\te',\om)V^{(1)}\cdots V^{(n)}\right>_{D_2}\vspace{3mm}\\
\dsty\hspace{2cm}\sim\int dq d\om q^{-1+\alpha'\om^2/4}\int d\te d\te' \left<V^i(\te,0)V^i(\te',0)V^{(1)}\cdots V^{(n)}\right>_{D_2}\vspace{3mm}\\
\dsty\hspace{4cm}\sim\left(p_1-p_2\right)^2\left<V^{(1)}\cdots V^{(n)}\right>_{D_2}\int dq \,d\om q^{-1+\alpha'\om^2/4}
\eer
$$
where we've taken into account that the operator $\int \del_n X(\te)d\te$ merely shifts the (straight) D-particle worldline and inserting it into any amplitude amounts to multiplication by the total transferred momentum $p_1-p_2$. If we now cut off the $dq\,d\om$-integral at $q=\eps$, the regularized value is proportional to $\sqrt{\log\eps}$. Since $q$ is introduced as a distance along the boundary of the worldsheet, the geometrical meaning of $\eps$ here is precisely the same as in the formula (\ref{D2}). Recalling that there is an additional power of $g_{st}$ in the annulus amplitude as compared to the disk amplitude with the same insertions,
we conclude that the modular integration divergence is proportional to
$$
g_{st}\left(p_1-p_2\right)^2\left<V^{(1)}\cdots V^{(n)}\right>_{D_2}\sqrt{\log\eps}\equiv g_{st}\left(p_1-p_2\right)^2 A_{D_2}^0\sqrt{\log\eps}
$$
To the order $1/M$ (i.e.~to the order $g_{st}$), we recognize precisely the same structure,
as in (\ref{D2}). We shall not work out here the actual coefficients that depend on a number
of conventions, but it is clear that the non-trivial dependencies on the momenta and the cut-off do in fact match, as it is necessary for the successful implementation of the Fischler-Susskind mechanism.

One cannot help but noticing that the version of the Fischler-Susskind mechanism that
we've just described bears a strong resemblance to the investigations of \cite{tafjord}.
Indeed, limiting themselves to the lowest order of the string coupling expansion, the authors of \cite{tafjord} have shown that, to that particular order, the consistency of the string S-matrix should be restored if the background of a straight D-particle worldline is augmented by an inclusion of the operator
$$
V_{TF}=\frac{p_1^\mu-p_2^\mu}{M}\int \del_n X_\mu(\te) X^0(\te)\,\Theta(X^0(\te)) d\te
$$
(where $\Theta(X^0)$ is a step function). Heuristically, such an operator corresponds
to the D-particle abruptly starting to move with the appropriate
recoil velocity at the moment $X^0=0$. (Once again, the operators $\int \del_n X^i(\te)d\te$
with $i$ being one of the Dirichlet directions shift the entire trajectory of the D-particle.)
The UV divergence induced by this operator is precisely what we've found in (\ref{D2}). Moreover, the divergent integral
$$
\int \frac{d\kappa}{\kappa^2}e^{-\kappa^2\log\eps}
$$
makes an actual appearance in the derivations of \cite{tafjord}. These parallels should
not be too surprising since, for example, the structure
$$
\int\limits_0^t d(\tilde t,\s)J^\mu(\s)d\tilde t\ds
$$
(entering the expression for the saddle point value of the D-particle trajectory $f^\mu(t)$) develops a discontinuity in its first derivative as the boundary
of the worldsheet shrinks to a single point.

The authors of \cite{tafjord} have remarked that the abruptness in the change of the
D-particle's velocity inherent to their scenario should probably be alleviated
at the higher orders of the recoil perturbation theory. In fact our present formalism
bypasses this problem from the very beginning. The summation over all possible D-particle trajectories automatically gives rise to the UV divergence required by the Fischler-Susskind mechanism, making it completely unnecessary to introduce the abrupt recoil by hand.

Is it possible to reach a deeper understanding of how the Fischler-Susskind mechanism
works in our present formalism beyond the lowest order of the recoil perturbation theory?
Extending our above analysis to the higher orders in $g_{st}$ is likely to require a
more systematic picture of both the general structure of the recoil perturbation theory and
the general structure of the divergencies in the modular integration. Yet, conceptually,
there doesn't seem to be anything puzzling about how the cancellations could actually occur.

Of course, the situation is much more intricate for the annihilation/pair-production case.
In that regime, the first non-trivial UV divergence will arise if we try to compute the
pre-exponential factor in the saddle point estimate (\ref{saddle}). There doesn't appear
to exist any meaningful expansion of this pre-exponential factor in any small parameter.
It is merely a determinant of some integral operator, whose divergent part needs to
be matched against the modular divergencies from the higher genus worldsheets.
Any cancellations in this setting will necessarily be rather subtle. In particular,
they may require certain relations between the values of the saddle point exponentials,
and therefore certain relations between the solutions to the saddle point equation
(\ref{tsaddle}) for worldsheets of different genera. We shall not pursue this line
of thought any further in this present investigation. Even though the considerations
related to the Fischler-Susskind mechanism are of a crucial importance for establishing
the consistency of the worldline formalism, they do not affect the estimates of the
saddle point exponential and, in particular, its coupling constant dependence, which is
the main result of this paper.

\section*{Emission of a large number of strings}

One may be somewhat worried by the fact that the annihilation amplitudes computed via the saddle point method
are suppressed much stronger in the limit $g_{st}\to 0$ than $\exp(-O(1/g_{st}))$,
which is generally
expected to be the order of the non-perturbative corrections in string theory \cite{shenker}.
The resolution of this apparent paradox is that the actual annihilation is most
likely to proceed into a large ($\sim 1/g_{st}$) number of quanta, whereas the
number of strings in the final state has been kept fixed as $g_{st}\to 0$ in
our preceding considerations. Indeed, as we could see in the above, the annihilation process 
was dominated by worldsheets of size $\sim 1/g_{st}$, and it would hardly be natural for
such extended configurations to decay into a small number of strings, a circumstance
responsible for the anomalously strong suppression of the amplitude (\ref{saddle}).

We shall now obtain some estimates of the amplitude for the D-particles to
annihilate into a large number of quanta, and show that those contributions are
in fact likely to saturate the bound of \cite{shenker}. The analysis of the previous
sections still applies, but the position of the saddle point (and hence the functional 
$\cal F$) now depends on $g_{st}$, and therefore extra care needs to be taken in
making statements about the coupling constant dependence of the amplitude.

It can be seen by inspecting the expression (\ref{saddle}) that, unless all but an
infinitesimal fraction of the final state strings are emitted from disconnected worldsheet
components of disk topology with exactly one vertex operator on each of them,
the $g_{st}^\chi$ factor itself will provide suppression of order $\exp(O(\ln(g_{st})/g_{st}))$.
The extent to which the dynamics (accounted for by the saddle point estimates) enhances
this suppression depends on the topology of each particular diagram.
When most of the final state strings are emitted from disks with a single vertex
operator, the non-perturbative suppression can only arise from the value of
the saddle point exponential. We shall now estimate the contribution from those diagrams.

When the worldsheet consists of many disconnected components each of which carries away
only a small fraction of the momentum, the independent displacements of those disconnected
components cause only a small change of the saddle point function. We therefore encounter
quasi-zero modes very similar to the ones described above (\ref{q0}). One way to account
for these quasi-zero modes is to introduce the integration over them explicitly into
the path integral by inserting $\de(\int_i t(\te)d\te-\bar t_i\int_i d\te)$ for each of the
disconnected components of the worldsheet (the $\te$ integral runs over the $i$th
component of the worldsheet, $\bar t_i$ thereby being the average position of its boundary).
The variational problem of (\ref{tsaddle}-\ref{zsaddle}) has then to be solved subject
to the constraints specified by the $\de$-functions. (This would introduce the corresponding Lagrange multipliers into the right-hand side of (\ref{tsaddle}), but they are inessential
to our subsequent considerations.)

In order to obtain constraints on the value of the amplitude, let us expand the 
${\cal P}(t)$ in (\ref{tsaddle}) in Taylor series around $\bar t_i$ (for each $i$). It is easy to see that unless more than $O(1/g_{st}^{1-\eps})$ (with $\eps$ arbitrarily small and positive) of the $\bar t_i$'s come into an interval of size $O(g_{st})$,
the solution to (\ref{tsaddle}) for $\te$ belonging to the $i$th disconnected
component is given by\footnote{We are essentially demanding the variation of 
$\cal P$ to be small over the extent of a single disconnected component of the worldsheet,
and then use the expression for the saddle point value of $t(\te)$ in the background of a straight worldline.}
$$
t(\te)=\bar t_i+\frac14\int \left[\del_n\del_{n'}D(\te,\te')\right]^{-1}\del_{n'}D(\te',\s')
{\cal P}^\mu(\bar t_i){\cal J}_\mu(\s')\ds' d\te'+O(g_{st}^{1+\eps})
$$
(The second term here is closely related to the saddle point which dominates
the recoil perturbation theory in the lowest order of $g_{st}$.) For the value of
the saddle point function (the exponential in the last line of (\ref{harness})), we obtain
$$
\ber{l}
\dsty-\pi\alpha'M^2\left(\int\limits_0^Tdt\,z(t){\cal P}^2(t)+\int d\te d\te' \del_n\del_{n'}D(\te,\te')
\int\limits_0^{t(\te)}{\cal P}(t)dt\int\limits_0^{t(\te')}{\cal P}(t')dt'\right)\vspace{3mm}\\
\dsty\hspace{1cm}=-2\pi i\,\alpha'M^2\sqrt{\wp^2}\,T\vspace{4mm}\\
\dsty+\frac{\pi\alpha'}4\sum\limits_i\int d\te d\te'\ds\ds' {\cal P}^\mu(\bar t_i)J_\mu(\s)\del_n D(\te,\s)\left[\del_n\del_{n'}D(\te,\te')\right]^{-1}\del_{n'} D(\te',\s')
{\cal P}^\nu(\bar t_i)J_\nu(\s')\vspace{4mm}\\
\dsty\hspace{3cm}+O(1/g_{st}^{1-\eps})
\eer
$$
The second line here merely provides for the correct pole structure (and disappears after
the application of the reduction formula). Each term in the sum from the third line is
of order $O(g_{st}^0)$ and negative ($\cal P$ is imaginary to the lowest order in
$g_{st}$). However, since there are $O(1/g_{st})$ terms, we should expect a
non-perturbative suppression of order $\exp(-O(1/g_{st}))$.

We still have to deal with the excluded region of integration over the $\bar t_i$'s,
namely, the configurations with more than $O(1/g_{st}^{1-\eps})$ of the $\bar t_i$'s in an interval of size $O(g_{st})$. The volume of this region is proportional to
$\exp(O(\ln(g_{st})/g_{st}^{1-\eps}))$, however, and this is
the amount of non-perturbative suppression provided regardlessly of the
value of the saddle point exponential. Barring any unfathomable cancellations, 
the sum of the contributions from the two regions would be of order $\exp(-O(1/g_{st}))$, in accordance with the general
estimates of the non-perturbative effects in string theory.

\section*{Bosonic D-particle decay}

It is a notable feature of the worldline formalism that most of the results obtained in the previous sections are immediately generalized to the case of bosonic D-particle decay that has received a large amount of attention within the tachyon condensation considerations.
Indeed, the natural proposal for the D-particle decay amplitude is
\beq
\left<k_1,\cdots,k_m|p\right>=\lim\limits_{p^2\to -M^2}
\int dx\,dx'\,e^{ipx}\,G(x,x'|\,k_1,\cdots,k_m) 
\label{decay}
\eeq
where $G(x,x')$ is as it has been defined in (\ref{master}). The meaning of this formula is that the endpoint of the D-particle worldline is unconstrained in the path integral, and the reduction formula is applied only to the starting point. The path integral thus describes a
``disappearing'' D-particle.

Given the formal similarities between the expressions (\ref{reduct}) and (\ref{decay}),
it is not surprising that the derivations presented above for the annihilation/pair-production case carry over to the decay process with a formal substitution $p_1=p$, $p_2=0$. One important
difference is that, for $p_1^2\ne p_2^2$ the quasi-zero mode described in the paragraph above (\ref{q0}) will be absent. The saddle point configuration of the worldsheet will be localized
near the endpoint of the D-particle worldline, as one would expect on general grounds. For that
reason, the factor of $T$ featured in (\ref{q0}) will not make an appearance for the case
of D-particle decay. This is reassuring, since it precisely replaces the double pole necessary
for the application of the reduction formula (\ref{reduct}) by a single pole necessary for the application of the reduction formula (\ref{decay}). The conclusions regarding the coupling constant dependence of the amplitude to the leading order of the saddle point approximation will remain intact, i.e. the exclusive decay amplitudes will be non-perturbatively suppressed
as $\exp[-O(1/g_{st}^2]$. 

It would be interesting to see to which extent the results obtained here are compatible
with the investigations of the closed string emission by a decaying D0-brane within the
tachyon condensation approach \cite{tachyon}. Unfortunately, the (non-self-consistent) neglect of the closed
string back reaction that underlies the
derivations of \cite{tachyon} makes it hard to judge which features of the emission spectrum
described there should be trusted.

\section*{Conclusions and speculations}

As the calculations of the previous sections reveal, the problem of D0-brane annihilation
appears in many ways much more tractable in our present context than the field-theoretical 
soliton-anti-soliton annihilation or, more generally, any other topological defect
annihilation process that the author is aware of. Indeed, the saddle point method
provides rigorous results regarding the coupling constant dependence for the annihilation
with a fixed number of final state quanta, and certain qualitative estimates can
be devised for final states whose multiplicity increases as $g_{st}$ goes to zero.
For a fixed number of final state closed strings (and for D0-brane pair production),
the amplitudes turn out to be suppressed as $\exp[-O(1/g_{st}^2)]$, much stronger than the
na\"\i ve expectation $\exp[-O(1/g_{st})]$.

For the case of closed string scattering off a D-particle, the worldline techniques developed
here offer a rather elegant construction of the expansion in string coupling constant that properly takes into account the D-particle recoil. In particular, we resolve the paradox related to the abruptness in the change of the D-particle velocity that made an unwelcome appearance in the previous attempts to address the same problem \cite{tafjord}.

It is a somewhat dissatisfying feature of our present formalism that only {\it exclusive}
amplitudes appear to be computationally accessible. Should one have had actual D-particles at one's disposal, exploring the refined properties of their annihilation (such as final state multiplicity distribution) would certainly provide a valuable insight into their nature.
Exclusive annihilation amplitudes would be {\it the} relevant theoretical prediction to guide this sort of experimentation. However, in most phenomenological applications (such as
cosmology), the most important quantity is the simplest characteristic of the annihilation process, namely, the total cross-section.

It is the non-trivial nature of the final state multiplicity distribution, i.e. the dominant role of the final states with a huge number of quanta, that makes the reconstruction of the
total cross-section from the exclusive annihilation amplitudes so hard. If one could overcome
this difficulty, it would be possible, for example, to examine in our setting the stretched D-string interconnection, a process that has so far only admitted a semi-quantitative
treatment \cite{d-string}. The generalization of our present formalism to the case of D-strings is not likely to pose much difficulty. The problem is that, in the cosmological
context, one is interested in the {\it total} interconnection probability. 

For the case of D-particle pair production, we have established a {\it very strong}
non-perturbative suppression (of order $\exp(-1/g_{st}^2)$). It would be interesting to
contemplate whether it may have anything to do with the exponential suppression of the
microscopic black hole pair production amplitudes argued by one of the sides in the dispute \cite{voloshin}.

Another amusing feature is that there does not appear to be a unitarity relation that
constrains the absolute normalization of the D-particle annihilation amplitude. It is
hard to judge at this point whether or not this represents an actual non-perturbative
ambiguity in string theory.

Since many qualitative features of the topological defect annihilation, such as
the non-perturbative suppression of the exclusive amplitudes and the high mean multiplicity
of the final state, appear to be rather general, it is natural to ask whether our
computation can shed any light on the field-theoretical soliton annihilation,
a process that has thus far defied an analytic treatment. The Christ and Lee
formalism of \cite{christ} provides a rather conspicuous link to our approach,
as it introduces explicitly the center-of-mass position of the soliton, a
worldline to be integrated over. It is certainly true that performing the path integral
over the fluctuations of the fields around the given trajectory of the soliton
is a much more difficult task than the worldsheet integrations pertinent to the D-particle case.
This circumstance is also likely to undermine the direct applicability of the saddle point method, since it seems to hinge upon the summation over the worldlines. Yet, one could try
to develop some intuition about the kinds of worldlines that contribute most, and
establish some bounds, say, on the coupling constant dependence. If such a program succeeds,
it would effectively provide a path-integral derivation of certain non-trivial properties
of the classical annihilation solution. Mathematically and physically, such a derivation
would present a fair amount of elegance.

\section*{Acknowledgements}

I would like to thank Steven Gubser and John Schwarz for the discussions pertaining to the general standing of this project, as well as Curtis Callan and John Preskill for providing the backgrounds on the field-theoretical soliton-anti-soliton annihilation. This work has been supported in part by the U.S. Dept. of Energy under Grant No. DE-FG03-92-ER40701.


\begin{thebibliography}{99}
\bibitem{rajaraman}
R.~Rajaraman, {\it Solitons and Instantons}, North-Holland
(1982), and references therein.
\bibitem{inspiration}
J.~Polchinski,
Phys.\ Rev.\ Lett.\  {\bf 75}, 4724 (1995).
\bibitem{hirano-kazama}
S.~Hirano and Y.~Kazama,
Nucl.\ Phys.\ B {\bf 499}, 495 (1997).
\bibitem{tachyon}A.~Sen,
  arXiv:hep-th/0410103, and references therein.
\bibitem{hashimoto}  A.~Hanany and K.~Hashimoto,
  arXiv:hep-th/0501031.
\bibitem{vilenkin}A.~Vilenkin, E.~P.~S.~Shellard, {\it Cosmic strings 
and other topological defects},\\ Cambridge University Press (1994).
\bibitem{preskill}
J.~P.~Preskill,
Phys.\ Rev.\ Lett.\  {\bf 43}, 1365 (1979).
\bibitem{voloshin}
S.~B.~Giddings and S.~Thomas,
Phys.\ Rev.\ D {\bf 65}, 056010 (2002);\\
M.~B.~Voloshin,
Phys.\ Lett.\ B {\bf 524}, 376 (2002).
\bibitem{1/N}E.~Witten,
Nucl.\ Phys.\ B {\bf 160}, 57 (1979).
\bibitem{drukier}
A.~K.~Drukier and S.~Nussinov,
Phys.\ Rev.\ Lett.\  {\bf 49}, 102 (1982).
\bibitem{ringwald}
A.~Ringwald,
Nucl.\ Phys.\ B {\bf 330}, 1 (1990);
O.~Espinosa,
Nucl.\ Phys.\ B {\bf 343}, 310 (1990).
\bibitem{christ}N.~H.~Christ and T.~D.~Lee,
Phys.\ Rev.\ D {\bf 12}, 1606 (1975).
\bibitem{fischler}W.~Fischler, S.~Paban and M.~Rozali,
Phys.\ Lett.\ B {\bf 381}, 62 (1996).
\bibitem{tafjord}V.~Periwal and O.~Tafjord,
Phys.\ Rev.\ D {\bf 54}, 3690 (1996).
\bibitem{volume}J.~Liu and J.~Polchinski,
Phys.\ Lett.\ B {\bf 203}, 39 (1988).
\bibitem{fradkin}E.~S.~Fradkin and A.~A.~Tseytlin,
Phys.\ Lett.\ B {\bf 163}, 123 (1985).
\bibitem{combinatorics}J.~Polchinski,
Phys.\ Rev.\ D {\bf 50}, 6041 (1994).
\bibitem{polyakov}A.~M.~Polyakov, {\it Gauge Fields and Strings}, 
Harwood Academic Publishers (1987),\\pp.\ 151-191.
\bibitem{ambjoern}J.~Ambj\o rn, B.~Durhuus, T.~Jonsson, {\it Quantum Geometry},\\ Cambridge University Press (1997).
\bibitem{gross}
D.~J.~Gross and P.~F.~Mende,
Phys.\ Lett.\ B {\bf 197}, 129 (1987).
\bibitem{fischler-susskind}W.~Fischler and L.~Susskind,
  Phys.\ Lett.\ B {\bf 171}, 383 (1986); Phys.\ Lett.\ B {\bf 173}, 262 (1986).
\bibitem{polchinski-fischler-susskind}J.~Polchinski,
  Nucl.\ Phys.\ B {\bf 307}, 61 (1988).
\bibitem{shenker}S.~H.~Shenker
in Cargese 1990, Proceedings, {\it Random surfaces and quantum gravity},\\pp.\ 191-200.
\bibitem{d-string}M.~G.~Jackson, N.~T.~Jones and J.~Polchinski,
  arXiv:hep-th/0405229.
\end{thebibliography}
\end{document}